\input phyzzx.tex
\tolerance=1000
\voffset=-0.0cm
\hoffset=0.7cm
\sequentialequations
\def\rl{\rightline}

\def\t1{{\tilde 1}}

\def\t{\theta}

\def\bh{Bekenstein--Hawking~}
\def\B{Bekenstein~}

\REF{\BEK}{J. Bekenstein, Lett. Nuov. Cimento {\bf 4} (1972) 737; Phys Rev. {\bf D7} (1973) 2333; Phys. Rev. {\bf D9} (1974) 3292.}
\REF{\HAW}{S. Hawking, Nature {\bf 248} (1974) 30; Comm. Math. Phys. {\bf 43} (1975) 199.}
\REF{\HOL}{G. 't Hooft, [arXiv:gr-qc/9310026]; L. Susskind, J. Math. Phys. {\bf 36} (1995) 6377, [arXiv:hep-th/9409089]; R. Bousso, Rev. Mod. Phys. {\bf 74} (2002) 825, [arXiv:hep-th/0203101].}
\REF{\ADS}{J. Maldacena, Adv. Theor. Math. Phys. {\bf 2} (1998) 231, [arXiv:hep-th/9711200]; S. Gubser, I. Klebanov and A. Polyakov, Phys. Lett. {\bf B428} (1998) 105,
[arXiv:hep-th/9802109]; E. Witten, Adv. Theor. Math. Phys. {\bf 2} (1998) 253, [arXiv:hep-th/9802150].}
\REF{\BEKE}{J. Bekenstein, Phys. Rev. {\bf D23} (1981) 287.}
\REF{\VER}{E. Verlinde, [arXiv:hep-th/0008140].}
\REF{\MASS}{V. Balasubramanian and P. Kraus, Comm. Math, Phys. {\bf 208} (1999) 413, [arXiv:hep-th/990212].}
\REF{\HP}{S. Hawking and D. Page, Comm. Math. Phys. {\bf 87} (1983) 577.}
\REF{\WITT}{E. Witten,  Adv. Theor. Math. Phys. {\bf 2} (1998) 505, [arXiv:hep-th/9803131].} 
\REF{\PRO}{J. Bekenstein, Found. Phys. {\bf 35} 2005) 1805, [arXiv:quant-ph/0404042].}
\REF{\LOWT}{D. Deutch, Phys. Rev. Lett. {\bf 48} (1982) 286.}
\REF{\TSOL}{J. Bekenstein, Phys. Rev. {\bf D27} (1983) 2262.}
\REF{\SPE}{G. W. Unruh and R. M. Wald, Phys. Rev. {\bf D25} (1982) 942.}
\REF{\STR}{A. Strominger, JHEP {\bf 0110} (2001) 034, [arXiv:hep-th/0106113].}
\REF{\EDI}{E. Halyo, JHEP {\bf 0203} (2002) 009, [arXiv:hep-th/0112093].}
\REF{\MAT}{T. Banks, W. Fischler, S. Shenker and L. Susskind, Phys. Rev. {\bf D55} (1997) 5112, [arXiv:hep-th/9610043].}
\REF{\FLA}{R. Bousso, Phys. Rev. Lett. {\bf 90} (2003) 121302, [arXiv:hep-th/0210295]; JHEP {\bf 0405} (2005) 050, [arXiv:hep-th/0402058].}
\REF{\GCEB}{E. F. Flanagan, D. Marolf and R. Wald, Phys. Rev. {\bf D62} (2000) 084035, [arXiv:hep-th/9908070].}

\singlespace
\rl{SU-ITP-09/24}
\pagenumber=0
\normalspace
\medskip
\bigskip
\titlestyle{\bf{On Entropy Bounds and Holography}}
\smallskip
\author{ Edi Halyo{\footnote*{e--mail address: halyo@.stanford.edu}}}
\smallskip
\centerline {Department of Physics} 
\centerline{Stanford University} 
\centerline {Stanford, CA 94305}
\smallskip
\vskip 2 cm
\titlestyle{\bf ABSTRACT}

We show that the holographic entropy bound for gravitational systems and the Bekenstein entropy bound for nongravitational systems are holographically related. Using the AdS/CFT 
correspondence, we find that the Bekenstein bound on the boundary is obtained from the holographic bound in the bulk by minimizing the boundary energy with respect the AdS radius or 
the cosmological constant. This relation may also ameliorate some problems associated with the Bekenstein bound.

\singlespace
\vskip 0.5cm
\endpage
\normalspace

\centerline{\bf 1. Introduction}
\medskip

It is well--known that the maximum entropy in a region of space is bounded by the entropy of the largest black hole that fits into that region, i.e. the holographic entropy bound. 
The Bekenstein--Hawking formula[\BEK,\HAW] shows that black hole entropy is given by the horizon surface area in Planck units.
$$S \leq S_{BH}={A_{hor} \over {4G}} \eqno(1)$$
As a result, in a strongly gravitational system tha maximum entropy in a given region is determined by the surface area of that region.
This relation is also the original inspiration for the holographic principle which requires that gravitational systems in a d--dimensional space 
be described by field theories on the ((d-1)--dimensional) boundary[\HOL]. The most striking example of holography is the celebrated AdS/CFT correspondence[\ADS].

On the other hand, there is another entropy bound that applies to nongravitational systems, i.e. the \B entropy bound[\BEKE] given by
$$S \leq S_{Bek}=2 \pi E R \eqno(2)$$
where $E$ and $R$ are the energy and size of the system respectively.

At first sight these two bounds seem completely unrelated since (1) applies to strongly gravitational systems whereas (2) applies to nongravitational ones.
(WE may naively say that the black hole entropy saturates $S_{Bek}$ as well as $S_{BH}$ but this would be wrong since we should not use $S_{Bek}$ for black 
holes which are strongly gravitating systems.) However, since one of the bounds is gravitational and the other is not they may naturally be related by holography. 
This proposal can be tested in the context of the AdS/CFT correspondence. The simplest possibility is that $S_{BH}$ in the bulk corresponds to $S_{Bek}$ on the boundary 
which clearly does not work since the two bounds are not saturated simultaneously. For a large AdS black hole with $R>L$, it is well--known that $S_{BH}=S_{bound}<S_{Bek}$ showing 
that this proposal does not work. 

In this paper, we show that there is indeed a holographic relation between (1) and (2) which is slightly more complicated than the naive one stated above. In order to appreciate 
this relation note that $S_{BH}$ is a bound for a given radius $R$ whereas $S_{Bek}$ is a bound for fixed $R$ and $E$. It is this extra dependence of $S_{Bek}$
on the energy that is crucial for our purposes. In addition, in the AdS/CFT correspondence, the entropy of an AdS black hole with a given radius $R$ is independent of the AdS 
radius,$L$ (or the cosmological constant), whereas varying $L$ changes the black hole mass and thus the energy of the boundary theory. Therefore, when we vary $L$,
$S_{BH}$ remains the same but $S_{Bek}$ changes due to the change in boundary energy. We show that $S_{Bek}$ (for fixed $R$ and varying $L$) is minimized for $L=R$, i.e. when the
black hole radius matches the AdS radius. Then, the \B bound is saturated.

Alternatively, we can view the \B bound not as an upper bound on entropy but as a lower bound on the energy of a system 
for a given size and entropy. Increasing $L$ (keeping $R$ and $S$ fixed) decreases the boundary energy.
Since the bound (2) can be seen as a minimum for the boundary energy what we need to do is to minimize the boundary energy $E$ with respect to
$L$ for a fixed $R$ and $S$. We find that the boundary energy is minimized for $L=R$ and for this value $S_{BH}=S_{Bek}$.
Therefore, for a black hole in the bulk with entropy $S_{BH}$, $S_{Bek}$ is the boundary entropy that corresponds to the minimized (with respect to AdS radius or the cosmological constant) 
boundary energy. Due to the holographic relation $L^{d-2}=G_d N^2$ minimizing the energy with respect to $L$ is equivalent to minimizing it 
with respect to $N$ or the central charge in the boundary theory. This holographic relation between the holographic and the \B entropy bounds can be considered a new derivation of the latter 
by using the former and holography in the context of AdS/CFT correspondence.

The \B bound is known to have a number of problems associated with it. First, there are no exactly pure field theories decoupled from gravity in nature so the bound can only be
seen as an approximate one for weakly gravitating systems. Second, $ER$ is not a Lorentz invariant so it is not clear how we should view this quantity in a
coordinate invariant way. Third, for very low temperature systems one needs to include the boundary effects which are ad hoc and system dependent in order not to violate the bound. 
Finally, there is the species problem which is due to the fact that the bound (2) is violated if the number of species of particles (or fields)
is greater than an $E$ and $R$ dependent bound. It seems that, if by using holography, we associate the \B bound with the entropy of an AdS black hole we may  
ameliorate these problems.


This paper is organized as follows. In Section 2 we show, in the context of the AdS/CFT correspondence, the relation between $S_{BH}$ in the bulk and $S_{Bek}$ on the boundary. 
In Section 3 we show how this holographic
relation between the two entropy bounds may solve the problems associated with the \B bound. Section 4 is a discussion of our results.

\bigskip
\centerline{\bf 2. The Relation Between \B and Holographic Entropy Bounds}
\medskip

The holographic entropy bound[\BEK,\HAW] gives the maximum entropy in a given volume of space (or $R$) for a gravitational system. On the other hand, the \B entropy bound gives the maximum
entropy for a given volume and energy in a pure field (or weakly gravitating) theory. Since the former bound holds for a strongly gravitational system whereas the latter for a 
nongravitational one, we may 
hope to relate them by holography[\HOL] which requires gravitational physics in $d$ dimensions to be described by a field theory on the $(d-1)$--dimensional boundary.
In this section we derive the relation between the \B and the \bh entropy bounds using the AdS/CFT correspondence[\ADS]. We mainly follow the description of ref. [\VER].  

Consider a system that saturates the \bh bound in the bulk, i.e. an AdS black hole.
A large black hole (with $R>L$) in $AdS_d$ is described by the metric
$$ds^2=-\left(1+{r^2\over L^2}-{{2G_d M} \over r^{d-3}}\right)dt^2 + \left(1+{r^2\over L^2}-{{2G_d M} \over r^{d-3}}\right)^{-1} dr^2+ r^2 d^2 \Omega_{d-2} \eqno(3)$$
where $L^{-2}=\Lambda$, $G_d$ is Newton's constant and $M$ is the mass parameter of the black hole. The black hole horizon is located at $r=R$ which solves
$$1+{R^2\over L^2}-{{2G_d M} \over R^{d-3}}=0 \eqno(4)$$
Then the black hole mass parameter $M$ is given by
$$M={R^{d-3} \over {2G_d}}+ {R^{d-1} \over {2G_L^2}} \eqno(5)$$
The energy of the black hole (as an excitation in AdS space) is related to the mass parameter by[\MASS]
$$E_{AdS}={(d-2) \over {8 \pi}}M \eqno(6)$$

The boundary theory is a CFT which lives on a sphere $S^{d-2}$ of radius $r>>L$[\VER]. The boundary metric is invariant under conformal transformations so we can rescale
the boundary coordinates to fix the radius of the $S^{d-2}$ to be $R$, i.e. the black hole radius. As a result, the energy of the boundary CFT is redshifted by
a factor $L/R$ compared to $E_{AdS}$. The CFT energy is then given by
$$E_{CFT}={{c(d-2)} \over {48 \pi}} {V \over L^{d-1}} \left(1+ {L^2 \over R^2}\right) \eqno(7)$$
where $V=R^{d-2}$ is the volume of the boundary and $c$ is the central charge defined by $c=3L^{d-2}/G_d$.
$E_{CFT}=E_E+E_C$ has two contributions. $E_E$ is the usual extensive term which gives the contribution of a $(d-2)$--dimensional gas. The second term, $E_C$ is subextensive
and gives the contribution of the Casimir energy of the CFT on $S^{d-2}$. The Hawking temperature of the black hole is given by
$$T_H={R \over {4 \pi L^2}}\left((d-1)+(d-3){L^2 \over R^2}\right) \eqno(8)$$
The temperarure of the boundary theory is related to $T_H$ by $T_{CFT}=(L/R)T_H$ due to the redshift. The entropy of the boundary theory can be obtained by using 
$T_{CFT}=(\partial E_{CFT}/ \partial S)_V$ and is
$$S_{CFT}={c \over {12}} {V \over L^{d-2}}={A \over {4G_d}} \eqno(9)$$
As expected the entropy of the boundary CFT matches the bulk entropy of the AdS black hole. Therefore a black hole in AdS space is described by a thermal state of the boundary CFT on
a sphere (including the Casimir effect).

The holographic entropy bound for the bulk which is saturated by the AdS black hole is the maximum entropy for a gravitational system. The \B entropy bound holds for nongravitational systems
such as the boundary CFT. Since the former bound is for a strongly gravitational system whereas the latter a nongravitational one we may hope to relate them by holography. 
The simplest possibility is to assume that both bounds are saturated simultaneously. However, it is easy to see that this idea does not work. The holographic bound is saturated 
for an AdS black hole 
as given by eq. (9). Clearly this does not satisfy the \B bound on the boundary. For large AdS black holes with $R>L$ we find that the boundary entropy is smaller than $S_{Bek}$ by a factor of
$[R/L+(L/R)]/2$.

This should not be surprising; in fact the naive idea above could not possibly work for the following reason. The bulk entropy for the black hole is independent of the cosmological
constant or $L$. If we vary $L$ keeping $R$ fixed, $S_{BH}$ remains constant whereas the bulk energy $E_{AdS}$ changes. As a result, $E_{CFT}$ and therefore $S_{Bek}$ changes. (Note
that the entropy of the boundary CFT which has to match the bulk entropy does not change.) Thus if this idea worked for a given $R$ and $L$ then the \B bound would be violated for
other values of $L$ with the same $R$. 

In order to find the correct relation between the entropy bounds we need to remember that the \B bound holds for fixed radius, $R$, and boundary energy, $E_{CFT}$. In fact, 
for our purposes it is better to consider the \B entropy bound as a (lower) bound for the energy of a nongravitating system for fixed $R$ and $S$.
$$E \geq {S \over {2 \pi R}} \eqno(10)$$
This simply means that the energy of the system (for fixed $R$ and $S$) is bounded from below. Therefore, in order to saturate the \B bound we need to minimize the boundary
energy $E_{CFT}$ keeping $R$ and $S$ fixed. Since $E_{CFT}$ is a function of the AdS radius, we can do this by varying $L$ i.e. the cosmological constant. In terms of the entropy, $E_{CFT}$ 
can be written as
$$E_{CFT}=S_{CFT}{{(d-2)} \over {4 \pi}} \left({1 \over L}+ {L \over R^2}\right) \eqno(11)$$
Keeping $R$ and $S$ fixed, we find that $E_{CFT}$ is minimized with respect to $L$ at $L=R$, i.e. when the black hole and AdS radii are equal. When $L=R$, the \B bound is saturated and
we see from eq. (9) that 
$$S_{CFT} = S_{Bek}={{2 \pi} \over {(d-2)}} E_{CFT} R \eqno(12)$$
This is the so--called normalized \B entropy which matches the usual \B entropy for $d=3$ and is stronger for $d>3$, in particular for $d=5$. The factor of $(d-2)$ clearly
arises from the relation between $E_{AdS}$ and the black hole mass parameter $M$ in eq. (6). It is a direct result of the fact that the bulk goemetry is AdS space with the
corresponding boundary CFT on $S^{d-2}$. We should then consider eq. (12) to be the \B bound rather than eq. (2) for the boundary CFTs realized through the AdS/CFT correspondence.
Thus, we find that the \B bound on the boundary is saturated when the holographic bound is saturated in the bulk (the AdS black hole) but only for a particular value of the AdS radius 
or the cosmological constant, i.e. the value that
minimizes the boundary energy (for fixed $R$). This seems to be the correct holographic relation between the holographic and \B entropy bounds.

We minimized the boundary energy with respect to $L$ which is a parameter of the bulk description. From the point of view of the boundary CFT, with fixed $G_d$ this is 
equivalent to minimizing with respect to $N$ due to the holographic relation $N^2=L^{d-2}/G_d$. Alternatively, this can be seen as minimizing the boundary energy with respect to the
central charge due to the relation $c=L^{d-2}/G_d=N^2$.

An alternative way of obtaining the same result is to notice that in general for $R>L$
$$S_{Bek}= {1 \over 2} S_{CFT} \left({R \over L} + {L \over R}\right) \eqno(13)$$
so that $S_{CFT} \leq S_{Bek}$. The bound is saturated for $R=L$ when $S_{Bek}$ is minimized with respect to $L$.

Starting with a large black hole ($R>L$) in AdS we can see what happens. For $R>L$ we have $S_{CFT}<S_{Bek}$. As we increase $L$ the bulk and boundary entropies remain constant 
but $E_{CFT}$ and therefore $S_{Bek}$ decrease. The minimum boundary energy and the $S_{Bek}$ are obtained for $L=R$ when $S_{CFT}=S_{Bek}$.

It is interesting that the AdS radius $L=R$ which saturates the \B bound is the same radius at which the Hawking--Page transition[\HP] in the bulk and the deconfinement transition on the
boundary[\WITT] take place. Also at $L=R$, the Casimir energy, $E_C$
(which is smaller than the extensive energy for $R>L$) on the boundary becomes equal to that of the gas $E_E$. It would be interesting to find out whether these facts are related to the
holographic relation between the two entropy bounds.. 

It is well--known that small AdS black holes with $R<L$ have negative specific heat, $C$, and are not stable. As a result, a small black hole 
in AdS space decays to a thermal gas due to the Hawking--Page phase transition[\HP]. Nevertheless we can consider small black holes to be metastable states if they have large masses 
(just as is customary for Schwarzschild black holes in flat space) and take a very long time to decay.
On the boundary a small black hole corresponds to a state of the CFT with a negative specific heat due to the dominance of the Casimir energy (with negative $C$) over that of the gas 
(with positive $C$). We may consider these unstable states in the CFT to be metastable if they correspond to small AdS black holes with large enough masses.
From eq. (9) we see that for small AdS black holes we again have $S_{CFT}<S_{Bek}$ since above $L=R$, $E_{CFT}$ and $S_{Bek}$ increase with increasing $L$ and fixed $R$ and $S$. Thus
$E_{CFT}$ and therefore of $S_{Bek}$ are minimized at $L=R$ from both above and below.

\bigskip
\centerline{\bf 3. The \B Entropy Bound and Holography}
\medskip

In the previous section we saw that the \B entropy bound on the boundary is obtained for an AdS black hole in the bulk (which saturates the holographic bound) when the boundary energy is
minimized with respect to the AdS radius or the cosmological constant. It is in this sense that the two entropy bounds (1) and (2) are related by holography. This relation may also 
constitute a new derivation of the \B bound using the holographic
bound and the AdS/CFT correspondence. (We remind that the original derivation of the \B bound also used the \bh entropy for black holes but of course not holography.)

Let us assume that we do not know about the \B entropy bound. Then, using the \bh entropy for an AdS black hole, the AdS/CFT correspondence and the fact that the bulk and boundary entropies
do not depend on the cosmological constant we would discover by the reasoning of the previous section that for a given $R$ and $S$, there is a minimum boundary energy, i.e. eq. (11).
We would observe that for this minimum energy case $S_{CFT}=S_{Bek}$ and for all other cases (with $L \not=R$) with higher boundary energy $S_{CFT}<S_{Bek}$ with $S_{Bek}$ given by eq. (12). 
In this derivation the bulk and boundary entropies remain constant as we vary $L$ but $S_{Bek}$ changes and attains its minimum at $L=R$. This result would be quite general 
(assuming holography) and
therefore we would conclude that there is a new entropy bound, i.e. the \B bound for nongravitating sytems such as the boundary CFT.

We now argue that the above relation between the two entropy bounds can be used to ameliorate some of the problems associated with the \B entropy bound[\PRO]. The idea is to use the (auxilary) 
gravitational bulk description dual to the nongraviational system under study. First, the \B entropy bound is usually 
used for weakly gravitating systems since in Nature there is no system completely decoupled from gravity. However, the boundary CFT which is dual to the gravitational bulk
theory is a pure field theory which is completely decoupled from gravity so the \B bound should apply to it precisely. Second, the product $ER$ that appears in the \B bound is not a Lorentz
invariant and therefore not well--defined in a relativistic theory. If we consider the dual bulk description through the AdS/CFT correspondence, we see that $E_{CFT}R=E_{AdS}L$. 
In the bulk, the mass of the black hole and the cosmological
constant are well-defined quantities. Therefore $ER$ on the boundary is well-defied through the holographic relation between the boundary and the bulk. 

Third, the \B bound seems to be violated for very low temperatures[\LOWT];
for example, for an ideal gas in a box of size $R$ this occurs when $T<< 1/R$. In this case we may either need to use the microcanonical ensemble and/or take into account  
boundary effects arising from the confinement of the gas[\TSOL] such as the Casimir effect. From eq. (8) we see that the temperature of the boundary CFT
cannot be smaller than a minimum value $\sim d/2 \pi L$ due to the contribution arising from the Casimir effect on the boundary $S^{(d-2)}$. Note that in eq. (8) the Casimir 
contribution to the temperature is automatically taken into account. 
Again, this makes sense only for large mass black holes which are metastable with a well--defined temperature. At the temperature $T=d/2 \pi L$ the deconfinement phase
transition takes place on the boundary so these black holes with $R<L$ correspond to supercooled states in the confined phase of the boundary theory. At very low temperatures, the
boundary theory is confined and has a very small entropy (O$(N^0)$) and should not violate the \B entropy bound.

Finally the \B bound suffers from the species problem, i.e. if the number of species, $N$, greater than a maximum value the \B bound is violated[\SPE].
In field theory, there is no upper limit for $N$ since the number of identical fields is a free parameter. However, the AdS/CFT correpondence imposes the holographc relation 
$N^2 G_{10}=L^8$ so $N$ is not a free variable. In fact, for given $R$ and $S$ in the bulk $G_{10}$ is fixed. Then we see that minimizing $E_{CFT}$ with respect to $L$ is equivalent to
minimizing it with respect to $N$ (or the central charge of the boundary theory). As we found above, this gives $L=R$ which also fixes the number of species to be $N^2=R^8/G_{10}$. 
Therefore, $N$ cannot be increased arbitrarily and the species problem is solved.

\bigskip
\centerline{\bf 4. Conclusions and Discussion}
\medskip

In this paper we showed that there is a holographic relation between the holographic and \B entropy bounds. This is quite natural since the former is a bound on strongly gravitational
systems whereas the latter holds for nongravitational ones and holography relates gravitational systems in the bulk to field theories on the boundary. We derived the relation, in the 
context of the
AdS/CFT correspondence, by considering a large black hole in the bulk and the corresponding thermal CFT on the boundary.  
Our derivation of this relation crucially depended on two observations. 
First, the bulk entropy of a large AdS black hole (and therefore the entropy of the boundary CFT) does not depend on the cosmological constant. Second, the \B entropy bound can also 
be viewed as a (lower) bound on the energy of a system
for a given size and entropy. Using the AdS/CFT correspondence, we showed that the \B entropy bound
on the boundary is obtained by minimizing the boundary energy with respect to the AdS radis or the bulk cosmological constant. 
We found that $E_{CFT}$ is minimized at $L=R$ at which
the \B bound in eq. (12) is saturated. The \B bound we derived is given by eq. (12) rather than the more common one in eq. (2) due to the factor $(d-2)$ which arises from the 
definition of the AdS bulk energy for a given black hole mass parameter $M$. 
Our result can be considered an alternative derivation of the \B entropy bound using the holographic entropy bound in the bulk and the AdS/CFT correspondence. In additon, 
we showed that the holographic relation between the two bounds ameliorates some of the problems associated with the \B bound such as the noninvariance of $ER$, violations due to 
very low temperatures and a large number of species.

It would be interesting to look for a similar relation between the two entropy bounds in de Sitter space. We can hope to do this by using the holographic dS/CFT correspondence[\STR].
A direct application of our method using the dS/CFT correspondence does not
work [\EDI]. This is not surprising since the dS/CFT correspondence has some fundamental problems and is quite different from the AdS/CFT correspondence even though superficially they look 
similar. Nevertheless, since we live in a universe with a positive cosmological constant, the de Sitter case is more relevant for Nature and deserves further study.
Similarly, we can try to find a holographic relation between the two entropy bounds in flat space, for example by using M(atrix) theory[\MAT]. Unfortunately, M(atrix) black holes are
described by CFT thermal states which do not saturate the \B entropy bound. Moreover, unlike the AdS case, in flat space we cannot vary the cosmological constant (which vanishes) 
so our method cannot possibly work. However, a derivaton of the \B bound in flat space[\FLA] which uses the generalized covariant entropy bound[\GCEB]. Unfortunately, it is hard to see 
a relation between our method and that used in [\FLA].

We were able to derive the \B bound by using the holographic bound in the bulk. It would be nice if we could do the opposite and derive the \bh entropy in the bulk from holography 
and the \B on the boundary. This does not
seem possible because we obtained the \B bound by minimizing the boundary energy. Knowing the minimum value of the boundary energy does not help us to derive 
the energy function for any $L$. In other words, each black hole in the bulk with a given $R$ (and $S$) that saturates the \bh bound corresponds to a different thermal state of the 
boundary CFT with $S_{CFT}<S_{Bek}$ but there is only one black hole (with $R=L$) that saturates the \B bound on the boundary.
In order to derive the \bh bound in the bulk from the \B bound we need to consider bulk states that do not saturate the \bh bound rather than black holes but on the boundary CFT and
these do not correspond to thermal states.

\bigskip
\centerline{\bf Acknowledgements}

I would like to thank the Stanford Institute for Theoretical Physics for hospitality.

\vfill

\refout

\end
\bye